\documentclass[12pt]{article}
\include{epsf}
\include{psfig}
\include{epsfig}
\usepackage{epsfig}
\begin{document}
\setcounter{page}{0}
\thispagestyle{empty}
\setlength{\parindent}{1.0em}
\begin{flushright}
GUTPA/04/12/02
\end{flushright}
\renewcommand{\thefootnote}{\fnsymbol{footnote}}
\begin{center}{\large{{\bf The Hierarchy Problem and an 
Exotic Bound State}}}\end{center}
\begin{center}{\bf C. D.
Froggatt}
\end{center}
\renewcommand{\thefootnote}{\arabic{footnote}}
\begin{center}{{\it  Department of Physics and Astronomy,
University of Glasgow,}\\{\it  Glasgow G12 8QQ, Scotland
}}\end{center}
\setcounter{footnote}{0}

\begin{abstract}
The Multiple Point Principle, according to which there
exist many vacuum states with the same energy density, is put
forward as a fine-tuning mechanism. By assuming the existence of
three degenerate vacua, we derive the hierarchical ratio between
the fundamental (Planck) and electroweak scales in the Standard
Model. In one of these phases, 6 top quarks and 6 anti-top quarks
bind so strongly by Higgs exchange as to become tachyonic and
form a condensate. The third degenerate vacuum is taken to have a
Higgs field expectation value of the order of the fundamental
scale.
\end{abstract}
\vspace {7cm}

{\small \it Plenary Talk at 10th International Symposium on Particles,
Strings and Cosmology (PASCOS04), North Eastern University, Boston, 
16-22 August 2004.}

\newpage

\section{Introduction}

The hierarchy problem refers to the long-standing puzzle of why the
electroweak scale is so very small compared to the fundamental scale
$\mu_{fundamental}$, which we shall identify with the Planck scale
$\mu_{Planck}$. In particular,
radiative corrections to the Standard Model (SM) Higgs mass diverge
quadratically with the SM cut-off scale $\Lambda$; this is the
so-called technical hierarchy problem. For example the
top quark loop contribution to the SM Higgs mass is given by:
 \begin{equation}
\delta M_H^2 = - \frac{3}{4\pi^2}g_t^2 \Lambda^2
\sim (300\ \hbox{GeV})^2
\left( \frac{\Lambda}{1\ \hbox{TeV}}\right)^2
\end{equation}
This leads to a fine-tuning problem for $\Lambda > 1$ TeV, when
the top quark loop contribution exceeds the physical SM Higgs
mass. For a SM cut-off at the Planck scale $\Lambda = \mu_{Planck}
\sim 10^{19}$ GeV, the quadratic divergencies have to be cancelled
to more than 30 decimals at each order in perturbation theory.

The most popular resolution of this technical hierarchy problem is
to introduce Supersymmetry or some other new physics (e.g.
technicolor or a little Higgs model) at the TeV scale. Although
SUSY stabilizes the hierarchy between the electroweak and Planck
(or GUT) scales, it does not explain why the hierarchy exists in
the first place. An alternative way to resolve the hierarchy
problem is to accept the necessity for fine-tuning and to
explicitly introduce a fine-tuning mechanism. The most well-known
example is the anthropic principle\cite{dine,arkani}. We shall
discuss here another fine-tuning mechanism: the Multiple Point
Principle. We shall apply it to the pure SM, with no new physics
below the Planck scale except presumably for a minor modification
at the see-saw scale to generate neutrino masses.

\section{Multiple Point Principle and the Large Scale Ratio}

According to the Multiple Point Principle\cite{brioni}, Nature
chooses the values of coupling constants in such a way as to
ensure the existence of {\em several} degenerate vacuum states,
each having approximately zero value for the cosmological
constant. This fine-tuning of the coupling constants is similar to
the fine-tuning of the intensive variables temperature and
pressure at the triple point of water, due to the co-existence of
the three degenerate phases: ice, water and vapour. The triple
point of water is an easily reproducible situation and occurs for
a wide range of the fixed extensive quantities: the volume, energy
and number of moles in the system.

We do not really know what is the dynamics underlying the Multiple
Point Principle, but it is natural to speculate that by analogy it
arises from the existence of fixed, but \underline{not} fine-tuned
extensive quantities in the Universe, such as
\begin{equation}
I_1 = \int d^4x \sqrt{g(x)} \quad \hbox{and} \quad I_2 = \int
\sqrt{g(x)} |\phi(x)|^2
\end{equation}
where $\phi(x)$ is the SM Higgs field. Such fixed extensive
quantities, having the form of reparameterisation invariant
integrals over space-time\cite{book} $I_i = \int d^4 x \sqrt{g(x)}
{\mathcal L}_i(x)$, can be imposed by inserting $\delta$-functions in
the Feynman path integral, similar to the energy fixing
$\delta$-function in the partition function for a microcanonical
ensemble in statistical mechanics\cite{fn2,corfu}. Then the
coefficient or coupling constant multiplying $I_i$ in the effective
action is constrained to lie in a very narrow range, analogous to
the inverse temperature in the canonical ensemble for a macroscopic
system with a fixed energy. The coupling constant acts as a
Lagrange multiplier, which
has to adjust itself to ensure that the extensive quantity $I_i$
takes on its correct fixed value. There is then a generic
possibility that, for a large range of values for $I_i$, the
Universe has to contain two or more degenerate phases in
different space-time regions. The imposition of a fixed value
for an extensive quantity is a non-local condition, which
seems to imply some mildly non-local physics, such as
wormholes or baby universes\cite{baby}, must underlie the multiple
point principle\cite{brioni,fn2,corfu}. However we emphasize that
the multiple point principle really has the status of a
postulated new principle.

We now wish to apply this fine-tuning principle to the problem of
the huge scale ratio between the Planck scale and the electroweak
scale: $\mu_{Planck}/\mu_{weak} \sim 10^{17}$. It is helpful at
this point to recall that another large scale ratio,
$\mu_{Planck}/\Lambda_{QCD} \sim 10^{20}$, is generally considered
to be a natural consequence of the SM renormalisation group
equation (RGE) for the QCD fine structure constant:
\begin{equation}
\frac{d}{d\mu}\left( \frac{1}{\alpha_3(\mu)} \right) =
\frac{7}{2\pi}
\end{equation}
Then taking $\alpha_3(\mu_{Planck}) \simeq 1/50$, which corresponds
to an order of unity value for the coupling constant at the Planck
scale $g_3(\mu_{Planck}) \simeq 1/2$, the RGE gives
$\mu_{Planck}/ \Lambda_{QCD} = \exp(2\pi/7\alpha_3(\mu_{Planck}))
\simeq \exp(45)$. A full understanding would of course require a
derivation of the value $g_3(\mu_{Planck}) \simeq 1/2$ from physics
beyond the SM, such as is done in the family replicated gauge
group model\cite{itepportoroz}.

Our proposed explanation for the large scale ratio
$\mu_{fundamental}/\mu_{weak}$ is similarly based on the use of the
RGE for the running top quark Yukawa coupling $g_t(\mu)$ in the SM:
\begin{equation}
%\beta_{g_t} =
\frac{dg_t}{d\ln\mu} =
 \frac{g_t}{16\pi^2}\left(\frac{9}{2}g_t^2 - 8g_3^2 -
 \frac{9}{4}g_2^2 - \frac{17}{12}g_1^2\right)
 \label{betatop}
\end{equation}
Here $g_3$, $g_2$ and $g_1$ are the $SU(3) \times SU(2) \times
U(1)$ running gauge coupling constants, which we shall consider as
given. The multiple point principle is used to fine-tune the
boundary values of $g_t(\mu)$ at {\it both} the fundamental and
weak scales, due to the existence of 3 degenerate SM vacua. Note
that we do \underline{not} use the physical top quark mass as an
input. These boundary values, $g_t(\mu_{fundamental})$ and
$g_t(\mu_{weak})$, then fix the amount of running needed from the
RGE (\ref{betatop}) and hence the required scale ratio
$\mu_{fundamental}/\mu_{weak}$.

\section{Two degenerate minima in the SM effective potential}

In order to fine-tune the value of $g_t(\mu_{fundamental})$ using
the multiple point principle, we postulate the
existence of a second degenerate vacuum\cite{fn2,fnt}, in which the
SM Higgs field $\phi$ has a vacuum expectation value of order
$\mu_{fundamental}$.
This requires that the renormalisation group improved effective
potential $V_{eff}(\phi)$ should have a second minimum near the
fundamental scale, where the potential should essentially vanish in order
to be degenerate with the usual electroweak scale minimum.

For large values of
the SM Higgs field $\phi \sim \mu_{fundamental} \gg \mu_{weak}$, the
renormalisation group improved effective potential is well approximated by
\begin{equation}
V_{eff}(\phi) \simeq \frac{1}{8}\lambda (\mu = |\phi | ) |\phi |^4
\end{equation}
and the degeneracy condition means that $\lambda(\mu_{fundamental})$
should vanish to high accuracy. The effective potential $V_{eff}$
must also have a minimum and so its derivative should vanish. Therefore
the vacuum degeneracy requirement means that the Higgs self-coupling
constant and its beta function should vanish near the fundamental
scale:
\begin{equation}
\lambda(\mu_{fundamental}) = \beta_{\lambda}(\mu_{fundamental}) =0
\label{boundary}
\end{equation}
This leads to the fine-tuning condition\cite{fn2}
\begin{equation}
\label{gt4}
g_t^4(\mu_{fundamental}) =
\frac{1}{48} \left(9g_2^4 + 6g_2^2g_1^2+3g_1^4 \right)
\end{equation}
relating the top quark Yukawa coupling $g_t(\mu)$ and the electroweak
gauge coupling constants $g_1(\mu)$ and $g_2(\mu)$ at
$\mu = \mu_{fundamental}$.
We must now input the experimental values of the electroweak gauge
coupling constants, which we evaluate at the Planck scale using the
SM renormalisation group equations, and obtain our prediction:
\begin{equation}
g_t(\mu_{fundamental}) \simeq 0.39.
\label{gtfund}
\end{equation}
However we note that this value of $g_t(\mu_{fundamental})$, determined
from the right hand side of Eq.~(\ref{gt4}), is rather insensitive to
the scale, varying by approximately $10\%$ between $\mu = 246$ GeV
and $\mu = 10^{19}$ GeV.

\section{Three degenerate vacua and the exotic bound state}
\label{nbs}

We now want to fine-tune the value of $g_t(\mu_{weak})$
using the multiple point principle. In order to achieve this,
it is necessary to have 2 degenerate
vacua which only deviate by their physics at the electroweak
scale. So what could the third degenerate SM vacuum be?
Different phases are most easily
obtained by having different amounts of some Bose-Einstein
condensate. We are therefore led to consider a condensate of a
bound state made out of some SM particles. We actually
propose\cite{itepportoroz,coral} a new exotic strongly
bound state made out of 6 top quarks and 6 anti-top
quarks -- a dodecaquark!
The reason that such a bound state was not considered
previously is that its binding is based on the collective
effect of attraction between several quarks due to Higgs
exchange.

The virtual exchange of the Higgs particle between
two quarks, two anti-quarks or a quark anti-quark pair yields an attractive
force in each case. For top quarks Higgs excahnge provides a strong force,
since we know phenomenologically that $g_t(\mu) \sim 1$. So let us
consider putting more and more $t$ and
$\overline{t}$ quarks together in the lowest energy relative S-wave
states. The Higgs exchange binding energy for the whole system becomes
proportional to the number of pairs of constituents, rather than to the
number of constituents. So {\em a priori}, by combining sufficiently many
constituents, the total binding energy could exceed the constituent
mass of the system! However we can put a maximum of $6t + 6\overline{t}$
quarks into the ground state S-wave. So we shall now estimate the binding
energy of such a 12 particle bound state.

As a first step we consider the binding energy $E_1$ of one of them to
the remaining 11 constituents treated as just one particle analogous
to the nucleus in the hydrogen atom. We assume that the radius of the
system turns out to be reasonably small, compared to the Compton wavelength
of the Higgs particle, and use the well-known Bohr formula for the binding
energy of a one-electron atom with atomic number $Z=11$ to obtain the crude
estimate:
\begin{equation}
E_1 = -\left(\frac{11g_t^2/2}{4\pi}\right)^2 \frac{11m_t}{24}.
\label{binding}
\end{equation}
Here $g_t$ is the top quark Yukawa coupling constant, in a normalisation
in which the top quark mass is given by $m_t = g_t \, 174$ GeV.

The non-relativistic binding energy $E_{binding}$ of the 12 particle system
is then obtained by multiplying by 12 and dividing by 2 to avoid
double-counting the pairwise binding contributions. This estimate only takes
account of the $t$-channel exchange of a Higgs particle between the
constituents. A simple estimate of the $u$-channel Higgs exchange
contribution\cite{itepportoroz} increases the binding energy by a further
factor of $(16/11)^2$, giving:
\begin{equation}
 E_{binding} = \left(\frac{11g_t^4}{\pi^2}\right)m_t
\label{binding2}
\end{equation}

We have so far neglected the attraction due to the exchange of gauge
particles. So let us estimate the main effect coming from gluon
exchange\cite{coral} with a QCD fine structure constant
$\alpha_s(M_Z) = g_s^2(M_Z)/4\pi = 0.118$,
corresponding to an effective gluon $t-\overline{t}$ coupling constant
squared of:
\begin{equation}
e_{tt}^2 = \frac{4}{3}g_s^2 \simeq \frac{4}{3} 1.5 \simeq 2.0
\end{equation}
For definiteness, consider a $t$ quark in the bound state; it interacts
with 6 $\overline{t}$ quarks and 5 $t$ quarks. The 6 $\overline{t}$
quarks form a colour singlet  and so their combined interaction with the
considered $t$ quark vanishes. On the other hand the 5 $t$ quarks combine
to form a colour anti-triplet, which together interact like a $\overline{t}$
quark with the considered $t$ quark. So the total gluon interaction of the
considered $t$ quark is the same as it would have with a single
$\overline{t}$ quark. In this case the $u$-channel gluon contribution
should equal that of the $t$-channel. Thus we should compare the
effective gluon coupling strength $2 \times e_{tt}^2 \simeq 2 \times 2 = 4$
with $(16/11) \times Zg_t^2/2 \simeq 16 \times 1.0/2 = 8$ from the Higgs
particle. This leads to an increase of $E_{binding}$ by a factor of
$(\frac{4+8}{8})^2 = (3/2)^2$, giving our final result:
\begin{equation}
 E_{binding} = \left(\frac{99g_t^4}{4\pi^2}\right)m_t
\label{binding3}
\end{equation}

We are now interested in the condition that this bound state should
become tachyonic, $m_{bound}^2 < 0$, in order that a new vacuum phase
could appear due to Bose-Einstein condensation. For this purpose we
consider a Taylor expansion in $g_t^2$ for the mass {\em squared} of the
bound state, crudely estimated from our non-relativistic binding
energy formula:
\begin{eqnarray}
m_{bound}^2 & = & \left(12m_t\right)^2 - 2\left(12
m_t\right)\times
E_{binding} + ...\\
& = & \left(12m_t\right)^2\left(1 -\frac{33}{8\pi^2}g_t^4 +
...\right)
\label{expansion}
\end{eqnarray}
Assuming that this expansion can, to first approximation, be trusted
even for large $g_t$, the condition $m_{bound}^2=0$ for the appearance
of the above phase transition with degenerate vacua becomes to
leading order:
\begin{equation}
\label{gtphase}
g_t|_{phase \ transition} =
\left(\frac{8\pi^2}{33}\right)^{1/4} \simeq 1.24
\end{equation}

We have of course neglected several effects, such as weak gauge
boson exchange, $s$-channel Higgs exchange and relativistic
corrections. In particular quantum fluctuations in the Higgs field
could have an important effect\cite{coral} in reducing
$g_t|_{phase \ transition}$ by up to a factor of $\sqrt{2}$. It is
therefore quite possible that the value of the top quark running
Yukawa coupling constant, predicted from our vacuum degeneracy
fine-tuning principle, could be in agreement with the experimental
value $g_t(\mu_{weak})_{exp} \approx 0.98 \pm 0.03$. Assuming this
to be the case, we can now estimate the fundamental to weak scale
ratio by using the leading order RGE (\ref{betatop}) for the top 
quark SM Yukawa coupling $g_t(\mu)$. It should be noticed that,
due to the relative smallness of the fine structure constants
$\alpha_i =g_i^2/4\pi$ and particularly of
$\alpha_3(\mu_{fundamental})$, the beta function $\beta_{g_t}$ for
the top quark Yukawa coupling constant, Eq.~(\ref{betatop}), is
numerically rather small at the fundamental scale. Hence we need
many $e$-foldings between the two scales, where
$g_t(\mu_{fundamental}) \simeq 0.39$ and $g_t(\mu_{weak}) \simeq
1.24$. The predicted scale ratio is quite sensitive to the input
value of $\alpha_3(\mu_{fundamental})$. When we input the value of
$\alpha_3 \simeq 1/54$ evaluated at the Planck scale, from the
phenomenological value of $\Lambda_{QCD}$ using the RGE for the SM
fine structure constants, we predict the scale ratio to be:
\begin{equation}
\mu_{fundamental}/\mu_{weak} \sim 10^{16} - 10^{20}
\label{ratio}
\end{equation}
\begin{figure}
\leavevmode \centerline{\epsfig{file=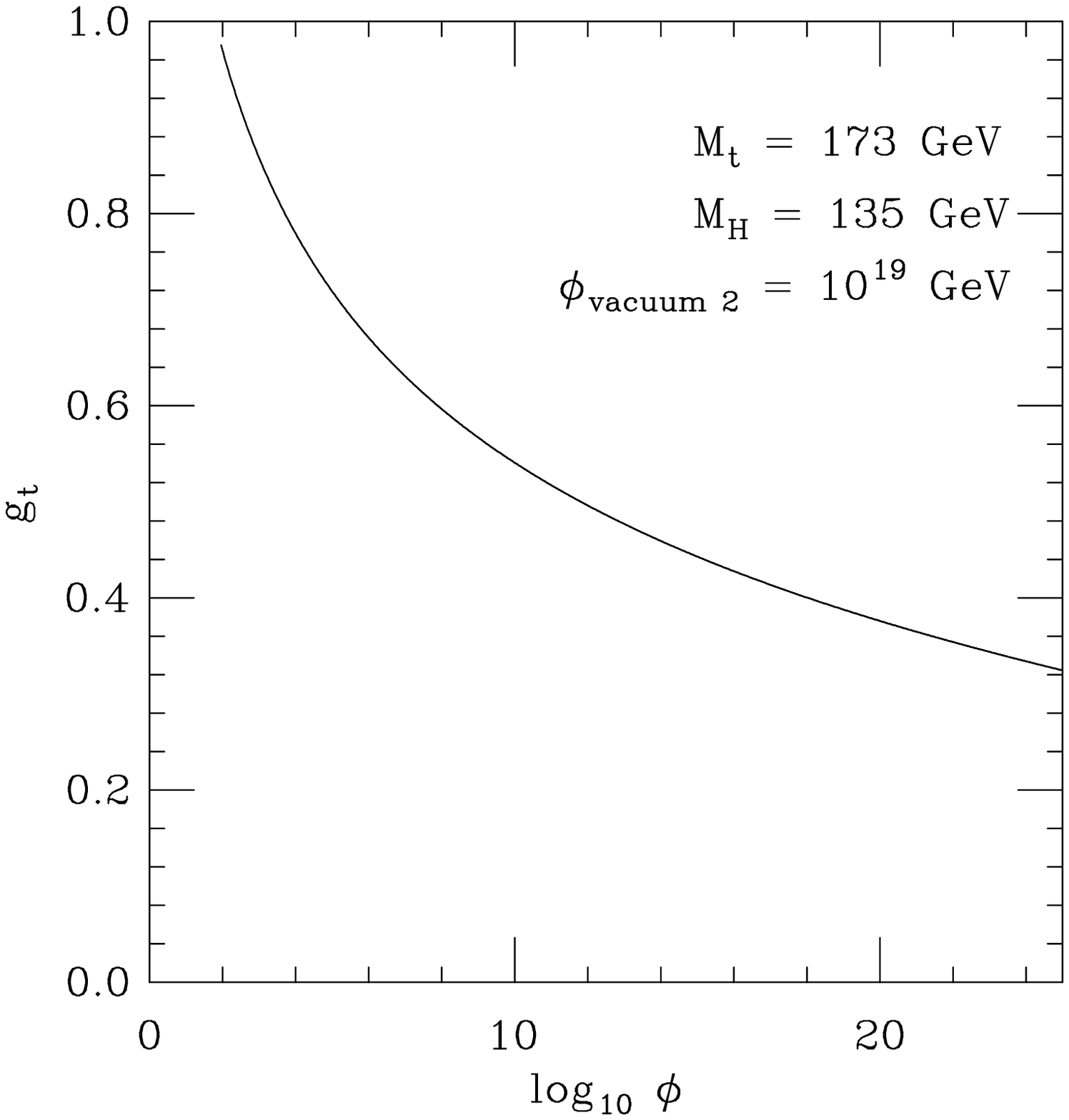,width=7.0cm}
\epsfig{file=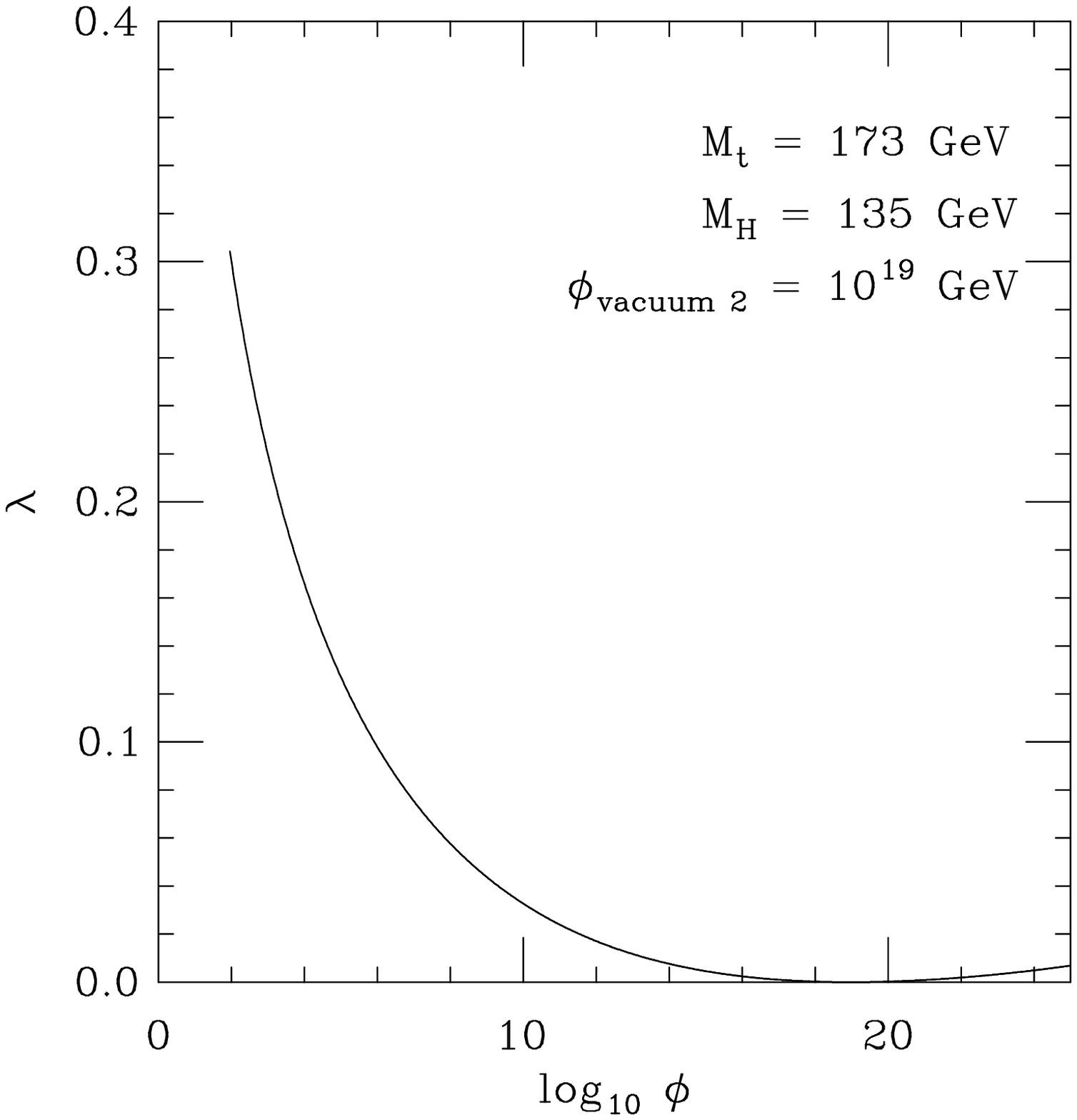,width=7.0cm} \hspace{-0.6cm}  }
\vspace{-0.6cm} \caption{Plots of $g_t$ and $\lambda$ as functions
of the scale of the Higgs field $\phi$ for degenerate vacua with
the second Higgs VEV at the Planck scale $\phi_{vacuum\;2}=10^{19}$
GeV. The second order SM renormalisation group equations are
formally applied up to a scale of $10^{25}$ GeV.}
\label{fig:lam19}
\end{figure}

The running of the top quark Yukawa coupling is shown in Figure
\ref{fig:lam19} as a function of $\log_{10}\phi$. We note that, as
can be seen from Eq.~(\ref{betatop}), the rate of logarithmic
running of $g_t(\mu)$ increases as the QCD gauge coupling constant
$g_3(\mu)$ increases. Hence the value of the weak scale is
naturally fine-tuned to be a few orders of magnitude above the QCD
scale. Using the RGE for the SM Higgs self-coupling
\begin{equation}
\frac{d\lambda}{d\ln \mu} = \frac{1}{16\pi^2} \left[ 12\lambda^2
+3(4g_t^2-3g_2^2-g_1^2)\lambda +\frac{9}{4}g_2^4 +
\frac{3}{2}g_2^2g_1^2 + \frac{3}{4}g_1^4 -12g_t^4 \right]
\end{equation}
and the boundary value at the fundamental scale,
Eq.~(\ref{boundary}), we can calculate the running of
$\lambda(\mu)$. The results are also shown in
Figure \ref{fig:lam19}. The value of $\lambda(\mu_{weak})$
obtained can be used to predict\cite{fn2} the SM Higgs mass
\begin{equation}
M_H=135 \pm 9 \ \hbox{GeV}
\end{equation}

\section{Properties of the exotic bound state}

Strictly speaking, it is {\em a priori} not obvious within our scenario
in which of the two degenerate electroweak scale vacua discussed in
Section \ref{nbs} we live. There is however good reason to believe
that we live in the usual Higgs phase without a condensate of
new bound states rather than in the one with such a condensate.
The point is that such a condensate is not invariant under the
$SU(2) \times U(1)$ electroweak gauge group and would contribute to
the squared masses of the $W^{\pm}$ and $Z^0$ gauge bosons. Although
these contributions are somewhat difficult to calculate, preliminary
calculations indicate that these contributions would make the
$\rho$-parameter deviate\cite{veltman} significantly from unity,
in contradiction with the precision electroweak data.

We expect the new bound state to be strongly bound and relatively
long lived in our vacuum; it could only decay into a channel in
which all 12 constituents disappeared together. The production
cross-section of such a particle would also be expected to be very
low, if it were just crudely related to the cross section for
producing 6 $t$ and 6 $\overline{t}$ quarks. It would typically
decay into 6 or more jets, but it would probably not be possible
to reconstruct the multi-jet decay vertex precisely enough to
detect its displacement from the bound state production vertex.
There would be a better chance of observing an effect, if we
optimistically assume that the mass of the bound state is close to
zero (i.e.~very light compared to $12m_t \approx 2$ TeV) even in
the phase in which we live. In this case the bound state obtained
by removing one of the 12 quarks would also be expected to be
light. These bound states with radii of order $1/m_t$ might then
be smaller than or similar in size to their Compton wavelengths
and so be well described by effective scalar and Dirac fields
respectively. The 6 $t$ + 6 $\overline{t}$ bound state would
couple only weakly to gluons whereas the 6$t$ + 5 $\overline{t}$
bound state would be a colour triplet. So the  6$t$ + 5
$\overline{t}$ bound state would be produced like a fourth
generation top quark\footnote{However there would be very little
mixing with the top quark, due to the small overlap of their wave
functions.} at the LHC. If these 11 constituent bound states were
pair produced, they would presumably decay into the lighter 12
constituent bound states with the emission of a $t$ and a
$\overline{t}$ quark. The 6$t$ + 6$\overline{t}$ bound states
would in turn decay into multi-jets, producing a spectacular event.

\section{Summary and Conclusion}

In this talk, we have put forward a scenario for how the huge
scale ratio between the fundamental scale $\mu_{fundamental}$ and
the electroweak scale $\mu_{weak}$ may come about in the pure SM.
We appeal to a fine-tuning postulate -- the Multiple Point
Principle -- according to which there are several different vacua,
in each of which the energy density (cosmological constant) is
very small. In fact our scenario requires a landscape of 3
degenerate SM vacua, in contrast to the $10^{1000}$ or so string
vacua\cite{dine,arkani}.

The existence of an exotic bound state of six top quarks and six
anti-top quarks is crucial to our scenario. Furthermore the
binding of this dodecaquark state, due mainly to Higgs particle
exchange, must be so strong that a condensate of such bound states
can form and make up a phase in which essentially tachyonic
bound states of this type fill the vacuum. The calculation of the
critical top quark Yukawa coupling $g_t|_{phase \ transition}$ for
which such a vacuum should appear involves no fundamentally new
physics. It is a very difficult SM calculation, but would provide
a clean test of the Multiple Point Principle as $g_t|_{phase \
transition}$ is predicted to equal the experimentally measured
value $g_t(\mu_{weak})_{exp}$. Within the accuracy of our crude
extrapolation (\ref{gtphase}) of the non-relativistic Bohr
formula, our Multiple Point Principle estimate of
$g_t(\mu_{weak})$ is in agreement with experiment.

In addition to the 2 degenerate electroweak scale vacua, we
postulate the existence of another degenerate vacuum in which the
SM Higgs field has a vacuum expectation value of order the
fundamental scale. We thereby obtain a prediction (\ref{gtfund})
for the value of $g_t(\mu_{fundamental})$ in terms of the
electroweak gauge coupling constants. The crucial point now is
that we need an appreciable running of $g_t(\mu)$, in order to
make its fine-tuned values at $\mu_{weak}$ and at
$\mu_{fundamental}$ compatible. That is to say we need a huge
scale ratio (\ref{ratio}), since the running is rather slow due to
the smallness of the SM fine structure constants $\alpha_i$ from
the renormalisation group point of view. It also naturally follows
that the electroweak scale lies within a few orders of magnitude
above $\Lambda_{QCD}$.

Finally we remark that, in our scenario, there are still quadratic
divergencies in the radiative corrections to the Higgs mass
squared at each order of perturbation theory. However the Multiple
Point Principle fine-tunes the bare parameters at each order of
perturbation theory, so as to ensure the equality of the energy
densities in the three different SM vacua. Indeed we obtain a
prediction for the Higgs mass: $M_H \sim 135$ GeV.

\section*{Acknowledgements}

I should like to thank Michael Vaughn and the Organising Committee
for their warm hospitality at this enjoyable and successful
symposium. I should also like to acknowledge helpful comments from
Paul Frampton, Peter Hansen and Michael Vaughn. Finally I must 
mention the many useful discussions with my collaborators 
Larisa Laperasvili and Holger Bech Nielsen.

\end{document}